\RequirePackage{rotating}
\documentclass[fleqn,usenatbib,useAMS]{mnras}

\usepackage{graphicx}	% Including figure files
\usepackage{amsmath}	% Advanced maths commands
\usepackage{array,multirow,makecell}
\defcitealias{C20}{Paper I}
\defcitealias{C22}{Paper III}
\usepackage[T1]{fontenc}
\usepackage{ae,aecompl}

\usepackage{newtxtext,newtxmath}
\title[MagES: SMC Outskirts]{The Magellanic Edges Survey -- IV. Complex tidal debris in the SMC outskirts}

\author[L. R. Cullinane et al.]{L. R. Cullinane$^{1,2}$\thanks{E-mail: lcullin4@jhu.edu (LRC)},
A. D. Mackey$^{2}$,
G. S. Da Costa$^{2}$,
S. E. Koposov$^{3,4}$, 
D. Erkal$^{5}$%,\newauthor
\\
$^{1}$Department of Physics and Astronomy, Johns Hopkins University, Baltimore, MD 21218, USA\\
$^{2}$Research School of Astronomy and Astrophysics, Australian National University, Canberra, ACT 2611, Australia\\
$^{3}$Institute for Astronomy, University of Edinburgh, Royal Observatory, Blackford Hill, Edinburgh EH9 3HJ, UK\\
$^{4}$Institute of Astronomy, University of Cambridge, Madingley Road, Cambridge CB3 0HA, UK\\ 
$^{5}$Department of Physics, University of Surrey, Guildford GU2 7XH, UK 
}

\date{Accepted XXX. Received YYY; in original form ZZZ}

\pubyear{2022}

\begin{document}
\label{firstpage}
\pagerange{\pageref{firstpage}--\pageref{lastpage}}
\maketitle

\begin{abstract}
	We use data from the Magellanic Edges Survey (MagES) in combination with Gaia EDR3 to study the extreme southern outskirts of the Small Magellanic Cloud (SMC), focussing on a field at the eastern end of a long arm-like structure which wraps around the southern periphery of the Large Magellanic Cloud (LMC). Unlike the remainder of this structure, which is thought to be comprised of perturbed LMC disk material, the aggregate properties of the field indicate a clear connection with the SMC. We find evidence for two stellar populations in the field: one having properties consistent with the outskirts of the main SMC body, and the other significantly perturbed. The perturbed population is on average \textasciitilde$0.2$~dex more metal-rich, and is located \textasciitilde7~kpc in front of the dominant population with a total space velocity relative to the SMC centre of \textasciitilde230~km~s$^{-1}$ broadly in the direction of the LMC. We speculate on possible origins for this perturbed population, the most plausible of which is that it comprises debris from the inner SMC that has been recently tidally stripped by interactions with the LMC. 
\end{abstract}

\begin{keywords}
	Magellanic Clouds -- galaxies: kinematics and dynamics -- galaxies: structure
\end{keywords}

%%%%%%%%%%%%%%%%%%%%%%%%%%%%%%%%%%%%%%%%%%%%%%%%%%

%%%%%%%%%%%%%%%%% BODY OF PAPER %%%%%%%%%%%%%%%%%%
\widowpenalty=0
\clubpenalty=0
\section{Introduction} \label{sec:intro}
The Large and Small Magellanic Clouds (LMC/SMC) are exemplary testbeds for studying the effects of tidal interactions on galaxy evolution. Dynamical models suggest the Clouds have likely repeatedly interacted with each other over several Gyr \citep[e.g.][]{beslaRoleDwarfGalaxy2012a, pardyModelsTidallyInduced2018a,C22}. The infall of the Clouds into the Milky Way gravitational potential has also influenced the system \citep[e.g.][]{C21,lucchiniMagellanicStream202021}. Each of these interactions leaves imprints on the star formation, structure, and kinematics of the Clouds, and at distances of \textasciitilde50~kpc and \textasciitilde60~kpc for the LMC and SMC respectively \citep{pietrzynskiDistanceLargeMagellanic2019,graczykDistanceDeterminationSmall2020a}, these signatures can be studied in detail to precisely constrain the history of the Magellanic system. 

As the smaller Magellanic galaxy, the SMC is particularly susceptible to perturbation through interactions with both the LMC and Milky Way. It is significantly extended along the line of sight (LOS), with depths of up to 20-30~kpc, particularly in its eastern regions \citep[e.g.][]{hatzidimitriouStellarPopulationsLargescale1989,nideverTidallyStrippedStellar2013,scowcroftCarnegieHubbleProgram2016,subramanianVMCSurveyXXIV2017, youssoufiStellarSubstructuresPeriphery2021}. Deep photometric studies \citep[e.g.][]{cioniVMCSurveyStrategy2011,nideverSMASHSurveyMAgellanic2017}, in addition to data from the Gaia satellite \citep{gaiacollaborationGaiaDataRelease2018b, gaiacollaborationGaiaEarlyData2021b}, have also revealed numerous extended stellar substructures surrounding the SMC \citep[e.g.][]{pieresStellarOverdensityAssociated2017a,mackeySubstructuresTidalDistortions2018, belokurovCloudsArms2019, massanaSMASHingLowSurface2020}. Full kinematic data are crucial in understanding the origins of these features, and by extension, the interaction history of the Clouds. Analysis of an eastern SMC foreground population \citep{omkumarGaiaViewStellar2020,jamesPresenceRedGiant2021} find it has motion distinct from the SMC body, and suggest it may have formed via tidal stripping of the SMC during its last pericentric passage around the LMC \textasciitilde$150$~Myr ago \citep{zivickProperMotionField2018b}. Studies of the inner ($\lesssim$5-6$^\circ$) SMC using 3D stellar kinematics \citep{deleoRevealingTidalScars2020a, zivickDecipheringKinematicStructure2021} reveal it is being torn apart, likely also due to this recent pericentric passage. However, LOS velocity information is scarce in the extreme SMC outskirts. 

The Magellanic Edges Survey \citep[MagES:][hereafter Paper I]{C20} is designed to fill this gap. MagES is a spectroscopic survey using the 2dF+AAOmega instrument on the Anglo-Australian Telescope \citep{lewisAngloAustralianObservatory2dF2002, sharpOptimalExtractionFibre2010}, providing LOS velocities and select abundance data for red clump (RC) and red giant branch (RGB) stars across the Magellanic periphery. In combination with astrometry from Gaia EDR3, it allows for the derivation of the 3D kinematics necessary to constrain Magellanic interactions. 

In this Letter, we present a detailed study of MagES field 3, located in the SMC outskirts near the end of a long arm-like structure south of the LMC. In the third MagES paper \citep[hereafter Paper III]{C22}, we suggested this field appears associated with the SMC; however, further analysis suggests it may contain two kinematically distinct populations. Intriguingly, despite its close proximity to other Magellanic structures -- including the likely-LMC-associated arm, mixed LMC-SMC debris \citep{chengKinematicalAnalysisSubstructure2022a}, and SMC foreground populations \citep[e.g.][]{jamesPresenceRedGiant2021} -- the characteristics of this field are unique. Here, we explore the properties of this field using full 3D kinematics, and suggest potential origins for the material within it. 

\section{Data} \label{sec:data}
\begin{figure}
	\centering
	\includegraphics[width=0.85\columnwidth]{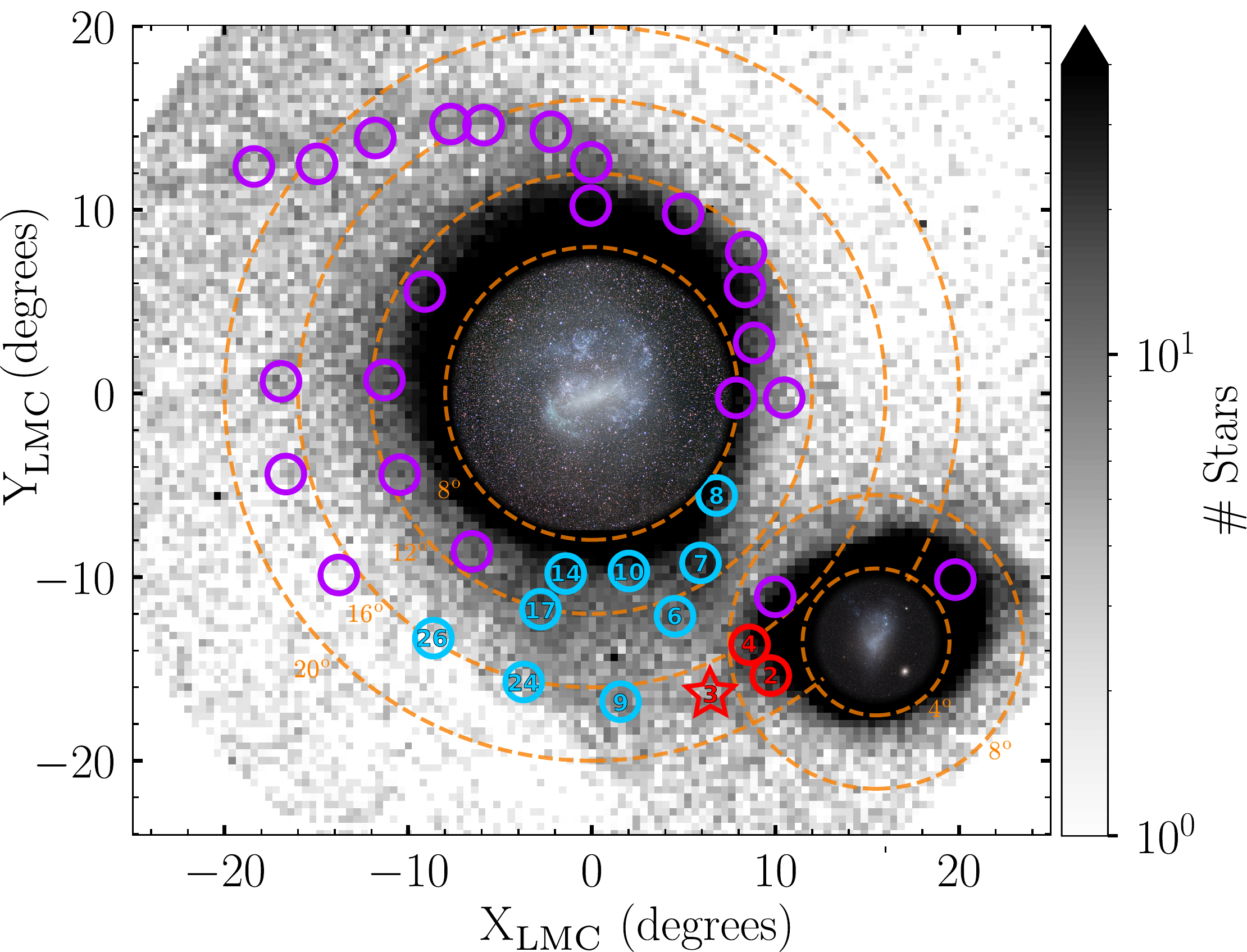}
	\caption{Location of MagES fields across the Magellanic periphery. The three SMC fields discussed in this letter are red, with field 3 starred. Reference fields in the outer LMC are blue, and fields not discussed here are purple. The background image shows the density of Magellanic red clump/RGB stars per square degree, selected from Gaia EDR3 as in \protect\citetalias{C22}. (X$_\text{LMC}$,Y$_\text{LMC}$) are coordinates in an orthographic projection centred on the LMC; north is up and east is to the left. Orange dashed circles mark angular separations of 8$^\circ$/12$^\circ$/16$^\circ$/20$^\circ$ from the LMC centre, and 4$^\circ$/8$^\circ$ from the SMC centre.}
	\label{fig:map}
\end{figure}

In addition to field 3, this letter discusses fields 2 and 4, which lie in the nearby southern SMC outskirts and are used as control fields. Field locations are shown in Fig.~\ref{fig:map}. Readers are referred to Papers I and III for a detailed description of the survey. We use the statistical framework described in \citetalias{C20} to probabilistically associate stars to either the Clouds, or one of several possible Milky Way contaminant populations, based on their kinematics. We typically use a single multi-dimensional Gaussian distribution -- characterised by the aggregate LOS velocity ($V_{\text{LOS}}$), proper motions ($\mu_\alpha,\mu_\delta$)\footnote{where $\mu_\alpha$ includes the usual $\cos(\delta)$ correction.}, and associated dispersions ($\sigma_{\text{LOS}},\sigma_\alpha,\sigma_\delta$) -- to describe the Magellanic population in a field. Nevertheless, our framework allows for additional components to be included (see \S\ref{sec:twopop}). Table~\ref{tab:aggprops} provides the location, inferred kinematic and photometric properties, and number of stars with probabilities $\geq$50\% of being associated with the Clouds, for each field discussed in this Letter. 

\begin{table*}
	%		\centering
	%		\captionsetup{textfont=footnotesize,labelfont={footnotesize,bf}}
	\caption{Properties for MagES fields analysed in this paper. Columns give the field centre position (RA, DEC in J2000.0); number of likely ($P_i\geq$50\%) Magellanic stars ($N_{\text{Mag}}$); on-sky SMC and LMC galactocentric radii ($R_{\text{SMC}}$/$R_{\text{LMC}}$); aggregate kinematic and photometric properties; and the field mean [Fe/H] with uncertainty 0.2~dex. All properties are calculated as described in \S\ref{sec:data}: we report the 68\% confidence interval as the $1\sigma$ uncertainty in each parameter.}
	\label{tab:aggprops}
	\tiny
	\setlength{\tabcolsep}{4pt}
	\begin{tabular}{lcccccllllllcllll} %@{\extracolsep{\fill}} %*{\textheight}
		\hline 
		\multicolumn{1}{>{\centering\arraybackslash}p{0.3cm}}{Field} & RA ($\alpha$) & DEC ($\delta$) & \multicolumn{1}{>{\centering\arraybackslash}p{0.4cm}}{$N_{\text{Mag}}$} & \multicolumn{1}{>{\centering\arraybackslash}p{0.5cm}}{$R_{\text{SMC}}$ \newline(deg)} & \multicolumn{1}{>{\centering\arraybackslash}p{0.5cm}}{$R_{\text{LMC}}$\newline(deg)} &  \multicolumn{1}{>{\centering\arraybackslash}p{0.8cm}}{$V_{\text{LOS}}$ \newline(km~s$^{-1}$)} & \multicolumn{1}{>{\centering\arraybackslash}p{0.8cm}}{$\sigma_{\text{LOS}}$\newline (km~s$^{-1}$)} & \multicolumn{1}{>{\centering\arraybackslash}p{0.8cm}}{$\mu_\alpha$\newline (mas~yr$^{-1}$)} & \multicolumn{1}{>{\centering\arraybackslash}p{0.8cm}}{$\sigma_\alpha$\newline (mas~yr$^{-1}$)} & \multicolumn{1}{>{\centering\arraybackslash}p{0.8cm}}{$\mu_\delta$\newline (mas~yr$^{-1}$)} & \multicolumn{1}{>{\centering\arraybackslash}p{0.8cm}}{$\sigma_\delta$\newline (mas~yr$^{-1}$)} & \multicolumn{1}{c} {$\left\langle(\text{BP}-\text{RP})_0\right\rangle$} & \multicolumn{1}{>{\centering\arraybackslash}p{0.7cm}}{$\sigma_{\text{(BP-RP)}_0}$} & \multicolumn{1}{c}{$\left\langle G_0\right\rangle$} & \multicolumn{1}{c}{$\sigma_{G_0}$} & \multicolumn{1}{c}{[Fe/H]} \\ \hline
		2 & 00 59 30.00 & $-$79 10 57.00 & 152 & 6.05 & 18.87 & $189.9\pm2.9$ & $34.3\pm2.2$ & $0.79\pm0.03$ & $0.25\pm0.03$ & $-1.30\pm0.02$ & $0.06\pm0.03$ & $0.91\pm0.03$ & $0.13\pm0.05$ & $19.00\pm0.05$ & $0.28\pm0.07$ & $-$1.6 \\ 
		4 & 01 45 11.00 & $-$79 15 22.00 & 152 & 6.86 & 16.74 & $185.5\pm2.2$ & $24.2\pm1.8$ & $1.14\pm0.03$ & $0.25\pm0.03$ & $-1.15\pm0.02$ & $0.12\pm0.03$ &  $0.92\pm0.02$ & $0.09\pm0.02$ & $18.96\pm0.03$ & $0.20\pm0.06$ & $-$1.6 \\
		3 & 01 20 00.00 & $-$82 30 00.00 & 68 & 9.45 & 18.07 & $185.4\pm4.1$ & $31.4\pm3.2$ & $1.41\pm0.07$ & $0.48\pm0.05$ & $-1.37\pm0.04$ & $0.27\pm0.05$ &  $0.93\pm0.03$ & $0.11\pm0.06$ & $18.77\pm0.07$ & $0.31\pm0.06$ & $-$1.4	\\ \hline
	\end{tabular}%* 
\end{table*}

For sufficiently bright ($G$>18) RGB stars\footnote{with six observed in field 3, but none in fields 2 and 4.}, MagES estimates [Fe/H] using the equivalent width of CaII triplet lines \citepalias[see][]{C20}. For fainter red clump stars, we stack spectra for likely ($P_i\geq$50\%) Magellanic stars to create a single “representative” spectrum used for these measurements; the resulting [Fe/H] estimates tend towards the mean metallicity of the given field. The mean red clump magnitude at the location of each field, used in this process, is calculated as in \citetalias{C22}.  All metallicity estimates have uncertainties of 0.2~dex. 

\section{Aggregate properties} \label{sec:onepop}
\begin{figure}
	\centering
	\hspace{0.7cm}\includegraphics[width=0.7\columnwidth]{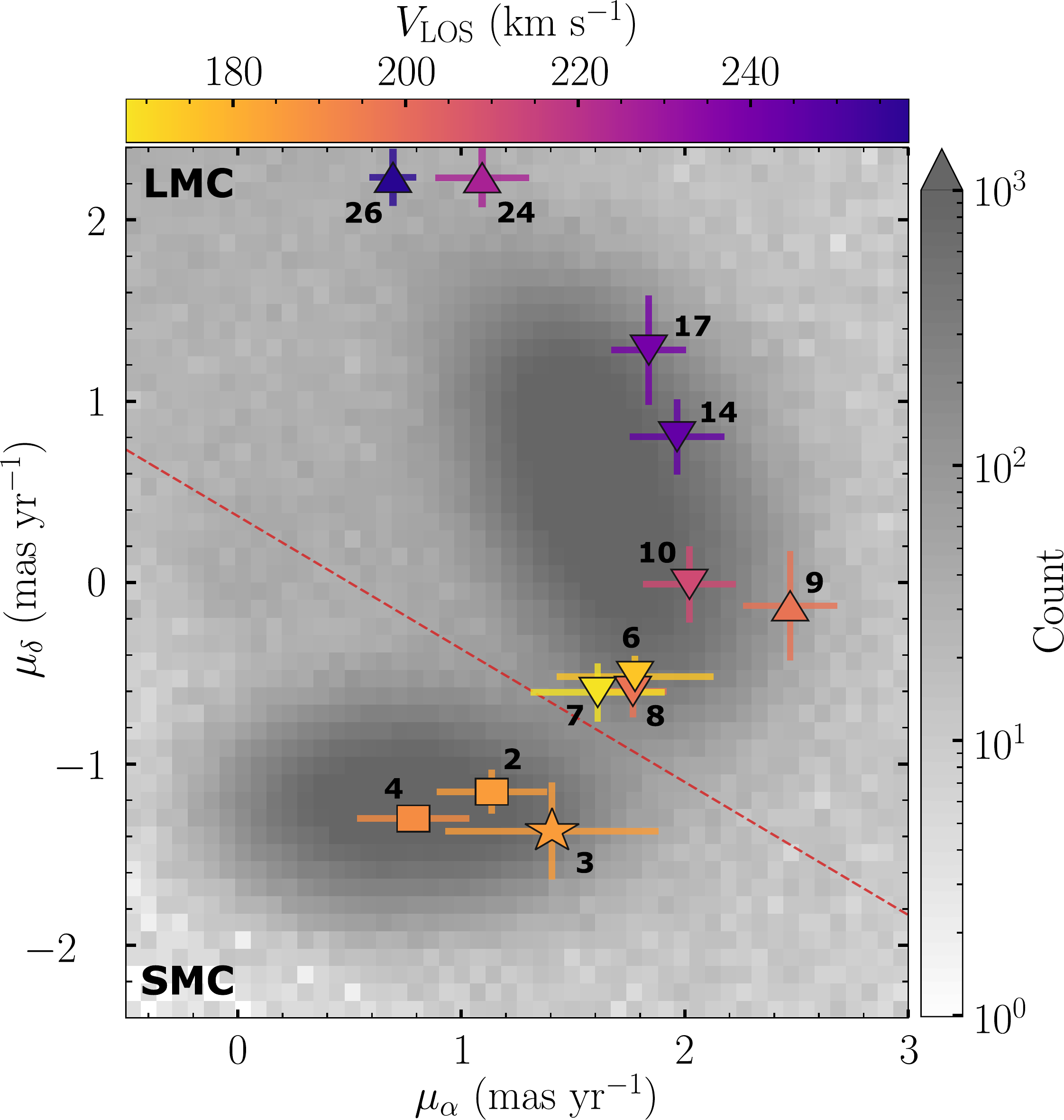}
	\includegraphics[width=0.64\columnwidth]{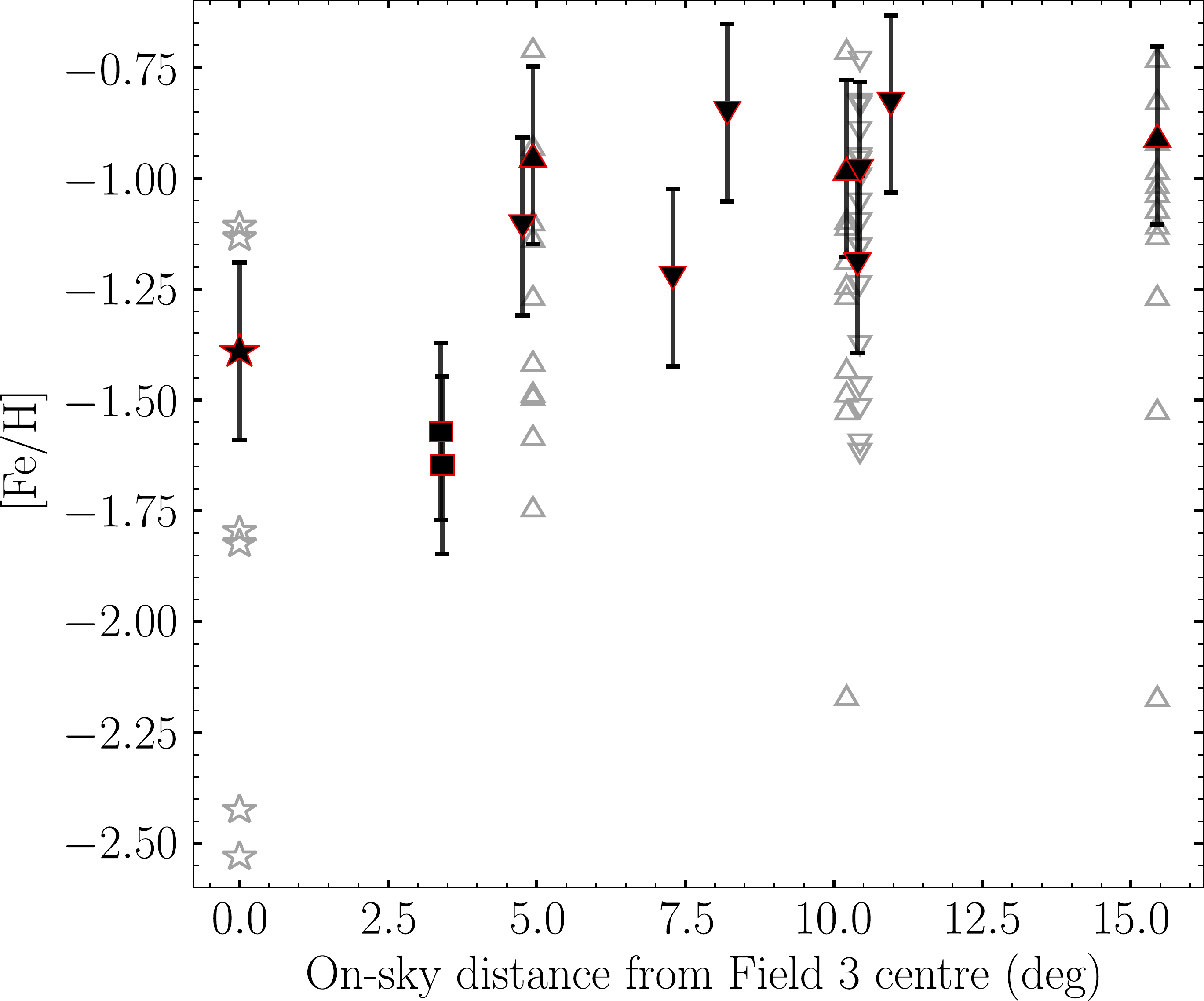}
	\caption{\textit{Top}: Aggregate proper motions for MagES fields, colour-coded by their aggregate LOS velocity. Symbols denote field 3 (starred point), fields 2/4  in the SMC outskirts (square points), LMC fields from \protect\citetalias{C22} along the southern arm-like structure (upwards-pointing triangles), and LMC southern disk and claw fields (downwards-pointing triangles, also from \protect\citetalias{C22}). These are overlaid on a 2D density plot of Gaia EDR3 proper motions for likely-Magellanic RC stars, selected as in \S\ref{sec:onepop}. The red dashed line shows an approximate boundary between LMC-like (upper right) and SMC-like (lower-left) proper motions. \textit{Bottom}: [Fe/H] estimates for MagES fields as a function of on-sky distance from field 3, with symbols as in the top panel. Solid points represent results from stacked spectra, which approximate the mean metallicity of the field. Open points represent [Fe/H] estimates for individual bright ($G$>18) RGB stars within fields where these are observed; associated uncertainties are omitted for clarity.}
	\label{fig:singleprops}
	\vspace{-10pt}
\end{figure}

Based on its location, field 3 could plausibly contain debris from both Clouds. Consequently, to obtain an initial estimate of the field’s composition, we compare its aggregate properties to those of nearby MagES fields comprised predominantly of material from only the LMC or SMC. The top panel of Fig.~\ref{fig:singleprops} presents the aggregate proper motions, colour-coded by the aggregate LOS velocity, for field 3 (starred point), fields 2 and 4  in the SMC outskirts, and several fields across the southern LMC outskirts from \citetalias{C22}. These are overlaid on a density plot of likely Magellanic RC stars selected from Gaia EDR3, with galactocentric radii $6^\circ$<$R_{\text{LMC}}$<$20^\circ$ and $2^\circ$<$R_{\text{SMC}}$<$10^\circ$. For this sample, we utilise the CMD selection box described in \S2.4 of \citetalias{C22}, imposing quality cuts \textsc{ruwe}<1.4 and $|C^*|$<$3\sigma_{C^*}$, a parallax cut $\varpi$<0.15 mas, and proper motion cuts $0.4$<$\mu_\alpha${(mas~yr$^{-1}$)}<$2.5,-1.6$<$\mu_\delta${(mas~yr$^{-1}$)}<$2.5$. The red dashed line marks an approximate boundary between LMC-like (upper right) and SMC-like (lower left) proper motions. The bottom panel of Fig.~\ref{fig:singleprops} presents [Fe/H] measurements for the same fields as a function of on-sky distance from the centre of field 3.%, using the same symbols as in the top panel. 

Fields 2 and 4 (square points in Fig.~\ref{fig:singleprops}) are clearly distinct from the LMC-associated reference fields. These fields sit firmly within the region of proper motion space associated with the SMC, and have mean metallicities \textasciitilde0.5~dex lower than fields dominated by LMC material: a difference broadly consistent with that expected given the relative masses of the Clouds. While the mean [Fe/H] values of these fields (\textasciitilde$-1.6$) are lower than literature metallicity measurements at smaller SMC radii \citep[\textasciitilde$-1$: e.g.][]{dobbieRedGiantsSmall2014a}, they do approximately follow the negative radial metallicity gradient in the SMC \citep[e.g.][]{cioniMetallicityGradientTracer2009,dobbieRedGiantsSmall2014a}, and also agree with photometric metallicity estimates in this region \citep{gradyMagellanicMayhemMetallicities2021}. We thus conclude that these fields are dominated by SMC material. 

Kinematically, field 3 is most similar to fields 2 and 4, with proper motions and a LOS velocity consistent with those of the SMC body. With a mean [Fe/H]\textasciitilde$-1.4$, field 3 is mildly more metal-rich than the nearby SMC fields (though still consistent within uncertainty), but somewhat more metal-poor than the average metallicity ([Fe/H] \textasciitilde$-1$) of the nearby LMC-dominated fields. This may indicate a mix of LMC and SMC material in the field; however, it is also consistent with expectations for stars originating in more central SMC regions.

\section{Two distinct SMC populations?} \label{sec:twopop}
\begin{figure*}
	\includegraphics[height=4.85cm]{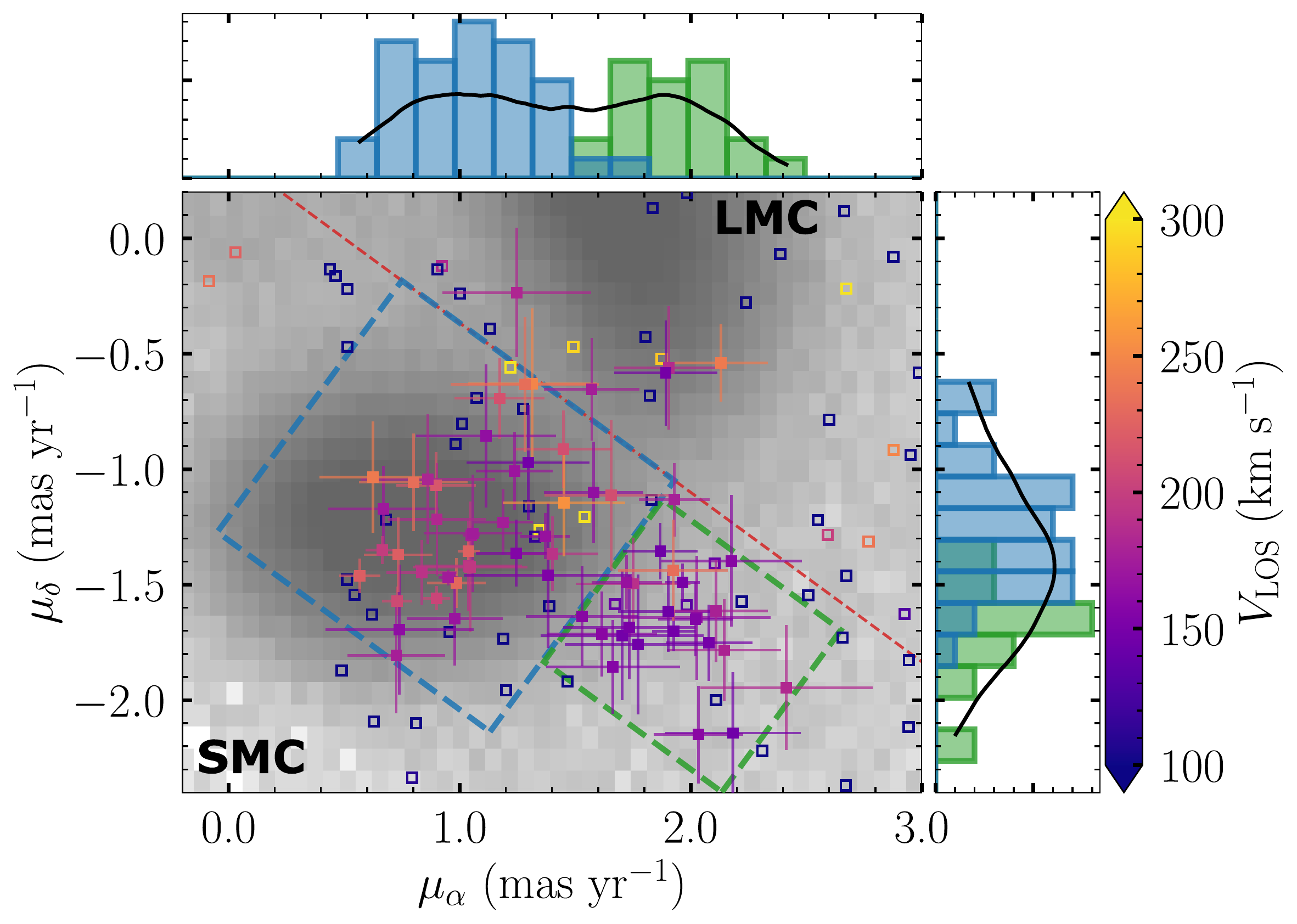}\hspace{0.15cm}
	\includegraphics[height=3.7cm]{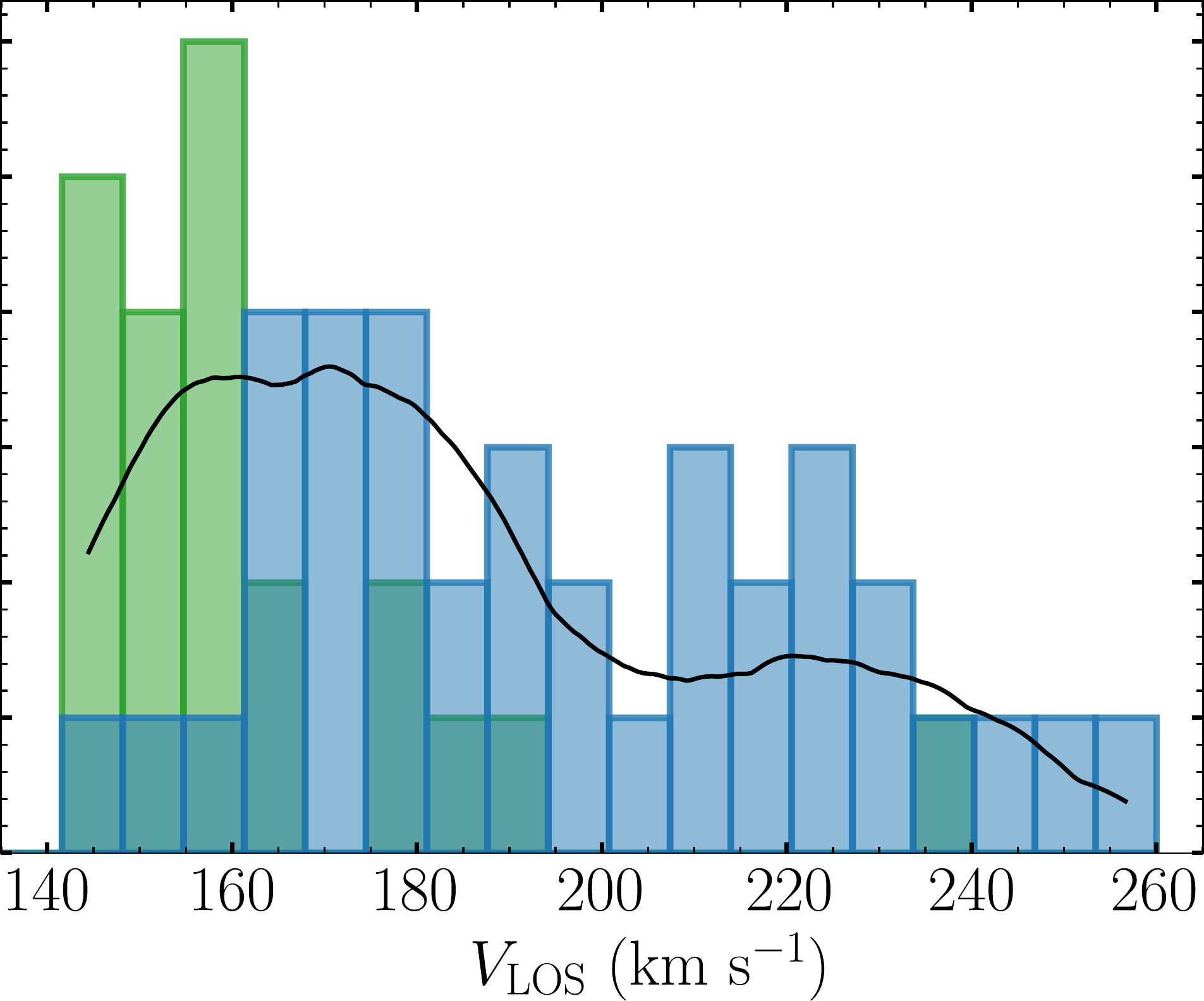}\hspace{0.15cm}
	\includegraphics[height=4.9cm]{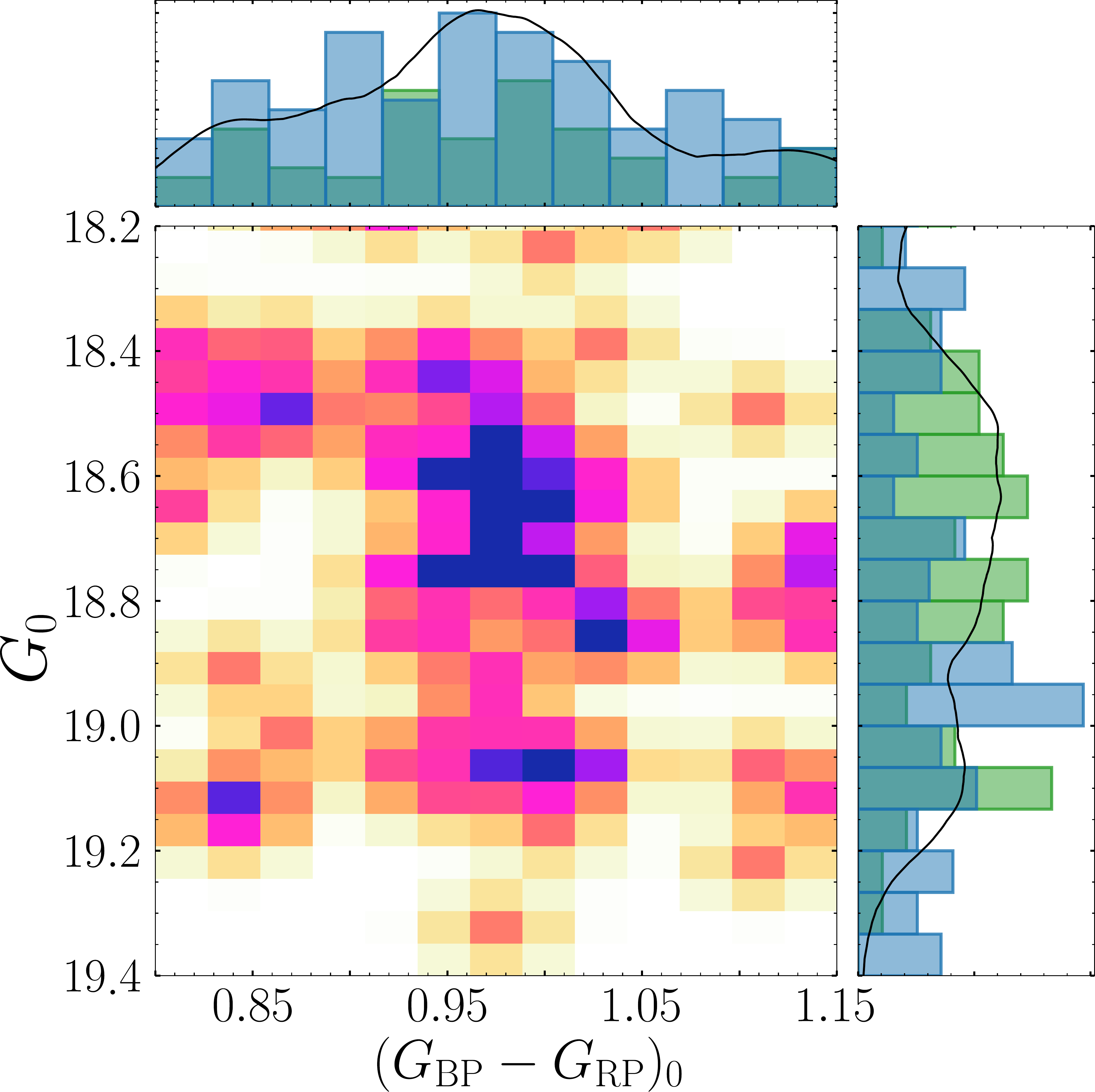}\hspace{0.05cm} 
	\raisebox{2.8ex}{\includegraphics[height=3.65cm]{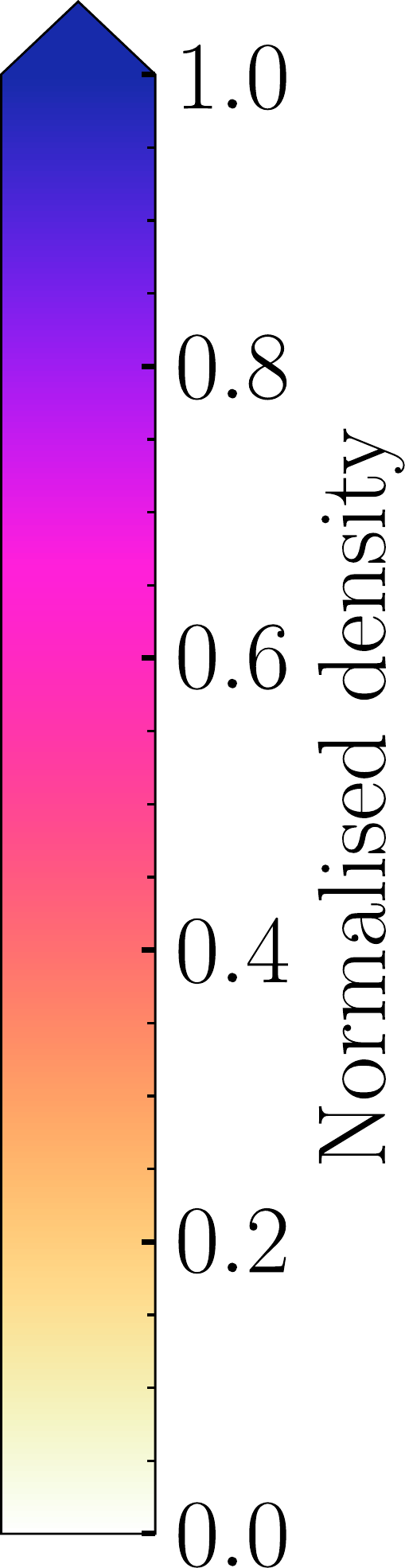}}
	
	\caption{\textit{Left:} Proper motion distribution of stars within field 3, colour-coded by their LOS velocity. Stars with probabilities <$50$\% of being associated with the Clouds based on fitting a single Magellanic population are represented as hollow points without associated uncertainties. The background image is as per Fig.~\ref{fig:singleprops}. Dashed lines indicate selection boxes used to define two subgroups in the field; the “bulk” population most consistent with the SMC body (blue), and the “offset” population (green). Top and side panels show proper motion histograms for the two subgroups (including only likely Magellanic members). 	Smooth curves overplotted in black are derived via kernel density estimation using an Epanechnikov kernel with bandwidth optimised using grid search cross-validation for all likely Magellanic stars. \textit{Centre:} LOS velocity histogram for the two subgroups. \textit{Right:} Hess diagram for stars within 1.5$^\circ$ of field 3, selected from Gaia as described below. Top and side panels show associated color and magnitude histograms for the two subgroups.} 
	\label{fig:2pops}
	\vspace{-7pt}
\end{figure*}

The preceding analysis assumes field 3 is comprised of a single population of stars. However, inspection of the LOS velocity and proper motion distributions of likely Magellanic members\footnote{i.e.\ with aggregate probabilities $\geq50$\% of being associated with the Clouds according to the single-component fit described in \S\ref{sec:data}; all probabilities quoted subsequently are also based on this fit.} in the field, as in Fig.~\ref{fig:2pops}, suggests the situation may be more complicated. Unlike other MagES fields, which have clear unimodal velocity distributions \citepalias{C20,C22}, the $\mu_\alpha$ distribution and, at lower significance, the LOS velocity distribution in field 3 both appear bimodal. Fig.~\ref{fig:2pops} thus hints at field 3 containing two populations of stars with distinct kinematics, highlighted by the dashed selection boxes. One set -- comprising \textasciitilde65\% of the Magellanic stars in the field -- is broadly consistent with the kinematics of the nearby SMC body\footnote{While the systemic SMC LOS velocity is \textasciitilde$150$~km~s$^{-1}$, the observed LOS velocity increases in a direction roughly towards field 3 due to the SMC’s disruption by the LMC \citep{deleoRevealingTidalScars2020a}.}, with total proper motion $\sqrt{\mu_\alpha^2+\mu_\delta^2}$\textasciitilde$1.6$~mas~yr$^{-1}$, and a median LOS velocity \textasciitilde$195$~km~s$^{-1}$. We henceforth refer to these stars as the “bulk” population. The putative second population has a larger total proper motion (\textasciitilde2.6~mas~yr$^{-1}$), but a lower median LOS velocity (\textasciitilde$160$~km~s$^{-1}$); we henceforth refer to these stars as the “offset” population. 

We test fitting these two potential populations using separate Gaussian distributions in addition to the MW foreground as in \S\ref{sec:data}. Comparing this with the single-component fit using the Bayesian Information Criterion \citep{schwarzEstimatingDimensionModel1978} shows there is insufficient evidence to prefer the two-component fit. However, in the restricted case of one- and two-component fits to only likely Magellanic members, we find the two-component fit {\it is} preferred. The relatively low number of stars in both populations and the moderate kinematic uncertainties for each individual star likely both contribute to the single-population solution being preferred when the background is included. 

Nonetheless, as the possibility of two populations is intriguing, we proceed with an investigation into the properties of the two groups of stars as defined in Fig.~\ref{fig:2pops}. For stars with Magellanic probability $\geq50$\% within each selection box, we present median 3D kinematics and dispersions (calculated as the standard deviation), in Table~\ref{tab:2p_props}. Associated uncertainties are determined via bootstrapping. These measurements confirm the kinematic differences between the bulk and offset populations indicated by the histograms in Fig.~\ref{fig:2pops}; varying the selection box limits does not change this conclusion. 

We additionally determine mean metallicity estimates for the two groups by stacking likely Magellanic RC star spectra as in \S\ref{sec:data}. This shows that the offset population is \textasciitilde0.2~dex more metal-rich than the bulk population. Photometric differences are also possible, but unfortunately there are insufficient likely Magellanic stars in each selection box to fit photometric RC properties using the methods described in \citetalias{C22}. Consequently, we expand our selection for this calculation only, taking stars from the Gaia sample described in \S\ref{sec:onepop}, within the selection boxes in Fig.~\ref{fig:2pops}, with on-sky radii $\leq1.5^\circ$ from field 3. Table~\ref{tab:2p_props} presents the resulting $G_0$ magnitude, $(G_{\text{BP}}-G_{\text{RP}})_0$ colour, and associated dispersions for the two groups, taking the 68\% confidence interval on these values as the associated uncertainty, and the right panel of Fig.~\ref{fig:2pops} presents a Hess diagram for the stars. 

\begin{table*}
	\caption{Properties of two subgroups of stars in field 3, discussed in \S\ref{sec:twopop}. Kinematic properties are calculated using MagES stars with probabilities $\geq50$\% of being associated with the Clouds, and photometric properties are calculated using Gaia-selected RC stars with on-sky radii $\leq1.5^\circ$ from the field centre as in \S\ref{sec:onepop}.}
	\label{tab:2p_props}

	\fontsize{7pt}{8.4pt} \selectfont
	\setlength{\tabcolsep}{4pt}
	\begin{tabular}{lcllllllcllll} 
		\hline 
		\multicolumn{1}{>{\centering\arraybackslash}p{0.55cm}}{Group} & \multicolumn{1}{>{\centering\arraybackslash}p{0.48cm}}{$N_{\text{Mag}}$} & \multicolumn{1}{>{\centering\arraybackslash}p{1.0cm}}{$V_{\text{LOS}}$ \newline(km~s$^{-1}$)} & \multicolumn{1}{>{\centering\arraybackslash}p{0.9cm}}{$\sigma_{\text{LOS}}$\newline (km~s$^{-1}$)} & \multicolumn{1}{>{\centering\arraybackslash}p{1.0cm}}{$\mu_\alpha$\newline (mas~yr$^{-1}$)} & \multicolumn{1}{>{\centering\arraybackslash}p{1.0cm}}{$\sigma_\alpha$\newline (mas~yr$^{-1}$)} & \multicolumn{1}{>{\centering\arraybackslash}p{1.1cm}}{$\mu_\delta$\newline (mas~yr$^{-1}$)} & \multicolumn{1}{>{\centering\arraybackslash}p{1.0cm}}{$\sigma_\delta$\newline (mas~yr$^{-1}$)} & $\left\langle(\text{BP}-\text{RP})_0\right\rangle$ & $\sigma_{\text{(BP-RP)}_0}$ & \multicolumn{1}{c}{$\left\langle G_0\right\rangle$} & \multicolumn{1}{c}{$\sigma_{G_0}$} & \multicolumn{1}{c}{[Fe/H]} \\ \hline
		Bulk & 37 & $195.2\pm7.5$ & $27.9\pm4.0$ & $1.05\pm0.08$ &  $0.28\pm0.04$ & $-1.25\pm0.08$ & $0.28\pm0.05$ & $0.92\pm0.04$ & $0.10\pm0.05$ & $19.00\pm0.06$ & $0.19\pm0.08$ & $-1.4$ \\
		Offset & 22 & $162.3\pm7.1$ & $19.7\pm8.5$ & $1.92\pm0.08$ & $0.21\pm0.05$ & $-1.68\pm0.07$ & $0.20\pm0.05$ & $0.98\pm0.01$ & $0.04\pm0.02$ & $18.74\pm0.07$ & $0.24\pm0.06$ & $-1.2$ \\ \hline
	\end{tabular}%* 
\end{table*}

These measurements show the offset population is mildly redder than the bulk population, likely a result of its \textasciitilde0.2~dex higher metallicity: in optical bands, metal-rich RC stars are intrinsically fainter and redder than those that are metal-poor \citep{girardiPopulationEffectsRed2001}. To quantify this, we use PARSEC isochrones \citep{bressanPARSECStellarTracks2012}\footnote{Accessed as version 3.4 at \url{http://stev.oapd.inaf.it/cmd}.} to test the predicted photometry variations in Gaia EDR3 passbands for the metallicities of the two populations. The isochrones, for an 11~Gyr population assuming default IMF parameters, predict that the measured metallicity difference results in colour variations of \textasciitilde0.05 mag, consistent with that observed. While age can also affect the photometric properties of the red clump \citep[e.g.][]{girardiPopulationEffectsRed2001}, the lack of younger ($\lesssim4$~Gyr) main sequence stars above an ancient (\textasciitilde11~Gyr) turnoff in deep DECam photometry of the field \citep{mackeySubstructuresTidalDistortions2018} implies a similar lack of young RC stars. 

The offset population is also \textasciitilde0.25 mag brighter than the bulk population. However this cannot be due to the difference in metallicity, as the more metal-rich offset population is nominally intrinsically fainter by \textasciitilde0.02~mag. This suggests the offset population is located substantially in front of the bulk population. A back-of-the-envelope calculation assuming the observed magnitude difference between the two populations is due entirely to distance effects, and that the bulk population has a distance of 60~kpc\footnote{The measured $\langle G_0\rangle$ for the bulk population is broadly consistent with isochrone predictions for a similar stellar population at this distance.}, implies the offset population is located \textasciitilde$7\pm2$~kpc in front of the bulk population. 

\section{Discussion} \label{sec:discussion}
The fact that splitting the Magellanic members in field 3 according to $\mu_\alpha$ leads to clear splits in metallicity, LOS velocity, and LOS distance, strongly supports the idea that two distinct populations are present. We can make several inferences regarding possible origins of this material based on the properties of these populations.

Given the similarity between the kinematic and photometric properties of the bulk population and the two nearby MagES SMC fields (see Table~\ref{tab:2p_props}), it seems likely that these stars comprise material from the local outskirts of the SMC. Further, models of debris produced in the SMC's recent disruption by the LMC in \citet{belokurovCloudsStreamsBridges2017} have similar kinematics to the bulk population in the vicinity of field 3, supporting this hypothesis. The origin of the offset population is, however, less clear; no material with similar kinematic properties is observed in the \citet{belokurovCloudsStreamsBridges2017} models. It is important to emphasise that the properties of field 3 cannot simply be explained by a single population spread across a large line-of-sight depth. This is in contrast to dual red clump features observed throughout much of the SMC \citep[e.g.][]{nideverTidallyStrippedStellar2013, subramanianVMCSurveyXXIV2017, youssoufiStellarSubstructuresPeriphery2021}, which do not show significant colour variations within individual fields, and which are well-described as the result of a distance spread much larger (>$12$~kpc) than that observed here. 

Although the offset population has some common characteristics with a foreground SMC substructure at galactocentric radii $\lesssim6^\circ$ \citep{omkumarGaiaViewStellar2020,jamesPresenceRedGiant2021} -- including larger proper motions, lower LOS velocities, and a closer distance than the main SMC body -- they are sufficiently different that we suggest a direct connection is unlikely. In particular, the proper motions of that foreground substructure ($\mu_\alpha,\mu_\delta$\textasciitilde1.1,-1.3~mas~yr$^{-1}$) are located entirely within our selection box for the bulk population. Further, the substructure is located \textasciitilde13~kpc in front of the main SMC population; nearly double the difference between the offset and bulk populations.

The comparatively high metallicity of the offset population relative to both the bulk population and nearby SMC fields indicates these stars are unlikely to be associated with the disrupted remains of an accreted Magellanic satellite. Since the offset population comprises only \textasciitilde1/3 of the likely Magellanic stars in field 3, this suggests any postulated disrupted satellite would have a low stellar mass, and hence be more metal-poor than the SMC \citep{kirbyUniversalStellarMassStellar2013}.

Another possibility is that the offset population is comprised of LMC stars: either associated with an LMC halo, or originating in the far outskirts of the LMC disk and subsequently perturbed during tidal interactions. For example, \citet{majewskiDiscoveryExtendedHalolike2008a} find a nominal LMC halo population having a similar mean metallicity ([Fe/H]\textasciitilde$-1.2$) to the offset population in fields at similar galactocentric radii, but located on almost the diametric opposite side of the LMC from field 3. The mean metallicity and red clump colour of the offset population are also consistent within uncertainty with nearby LMC-disk-associated MagES fields (cf. \citetalias{C22}), indicating they may be comprised of similar stellar populations.
However, the offset population kinematics are significantly different than expected in these scenarios. The much lower LOS velocity of the offset population (\textasciitilde160~km~s$^{-1}$) compared to the systemic velocity of the LMC \citep[\textasciitilde260~km~s$^{-1}$:][]{vandermarelThirdEpochMagellanicCloud2014} is inconsistent with that predicted for either the LMC disk or halo at the position of field 3. The offset population also has a total space velocity relative to the LMC centre (\textasciitilde140~km~s$^{-1}$) approximately double that of the most perturbed MagES LMC substructure fields (cf.\ \citetalias{C22}), rendering an association with the LMC unlikely. 

\begin{figure}
	\centering
	\includegraphics[width=0.75\columnwidth]{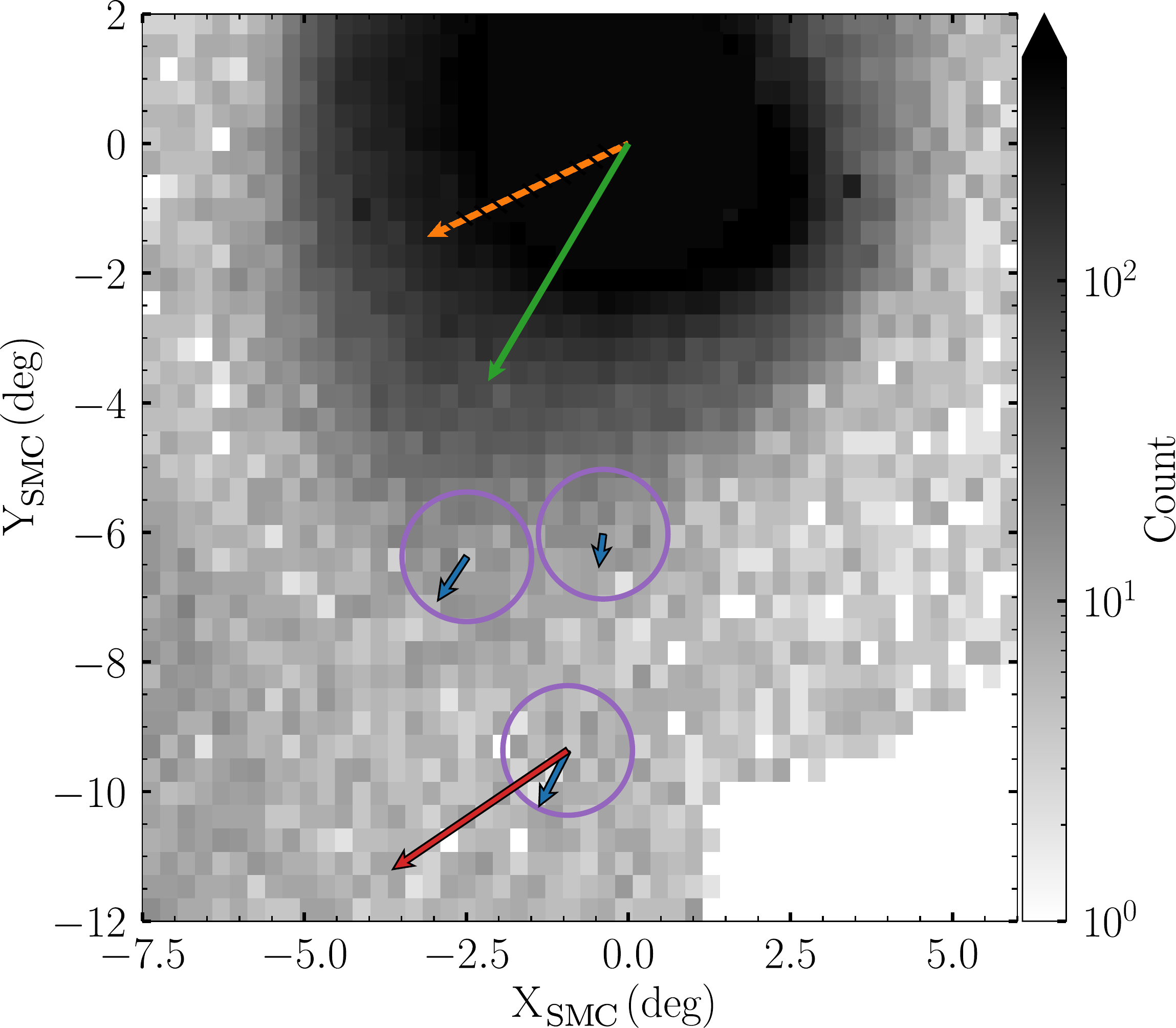}
	\caption{Projected proper motions for MagES fields (purple circles) in the SMC outskirts relative to the SMC centre. Blue arrows show motions for fields 2, 4, and the bulk population in field 3; the red arrow shows the motion of the offset population in field 3. The green arrow indicates the systemic proper motion of the SMC, with the dashed orange arrow indicating the direction of the LMC. The background image shows the stellar density for the same sample of Magellanic stars as in the top panel of Fig.~\ref{fig:singleprops}. The orientation of this figure is rotated \textasciitilde90$^\circ$ clockwise relative to Fig.~\ref{fig:map}.}
	\label{fig:vec}
	\vspace{-7pt}
\end{figure}

A more plausible origin is that the offset population comprises material originating at substantially smaller SMC radii, which has been perturbed and driven outwards. Given the mean metallicity in the inner SMC is approx.\ $-1$ \citep[e.g.][]{dobbieRedGiantsSmall2014a}, the metallicity of the offset population (\textasciitilde$-1.2$) is broadly consistent with this scenario. As seen in Fig.~\ref{fig:vec}, which presents the projected velocities of the two populations relative to the SMC centre, the velocity of the offset population is broadly in the direction of the LMC -- much more so than for the bulk population and nearby SMC fields. The large space velocity of the offset population (\textasciitilde230~km~s$^{-1}$) naively implies a time of just over 50~Myr to cover the distance between the SMC centre and the position of field 3, suggesting the material may be recently perturbed. The SMC pericentric passage around the LMC \textasciitilde150~Myr ago \citep{zivickProperMotionField2018b}, thought to be a near head-on collision at a distance $\lesssim$10~kpc from the LMC centre \citep{choiRecentLMCSMC2022a}, thus seems a likely contender for the source of this perturbation. 

We also note there are a few stars (<$10$) in fields 2 and 4 with velocities within $1\sigma$ of the median offset population kinematics, which may have similar origins. Unfortunately, there are too few such stars to reliably check the consistency of metallicity or photometric properties. %to test if these are similarly consistent. 
Notably, however, there are no stars in the nearby LMC-associated fields with kinematics matching the offset group, adding weight to our hypothesis that this population is linked to the SMC. 

In summary, we have discussed the complex kinematics of MagES field 3, located in the extreme southern outskirts of the SMC. While the aggregate properties of the field are consistent with an SMC origin, there are indications that two distinct sets of stars may be present: a “bulk” population with similar characteristics to the outskirts of the main SMC body, and an “offset” population with distinct photometric, kinematic, and chemical properties. We discuss several possible origins for this offset population, concluding that it is most plausibly comprised of debris from the inner SMC recently perturbed by the LMC. Our analysis illustrates the power of 3D kinematics in identifying structures otherwise invisible due to projection, and which -- in conjunction with future detailed modelling -- potentially hold great power for constraining the details of Magellanic interactions.

\section*{Acknowledgements}

This work has made use of data from the European Space Agency (ESA) mission {\it \textit{Gaia}} (\url{https://www.cosmos.esa.int/gaia}), processed by the {\it \textit{Gaia}} Data Processing and Analysis Consortium (DPAC, \url{https://www.cosmos.esa.int/web/gaia/dpac/consortium}). Funding for the DPAC has been provided by national institutions, in particular the institutions participating in the {\textit{Gaia}} Multilateral Agreement. Based on data acquired at the Anglo-Australian Observatory. We acknowledge the traditional owners of the land on which the AAT stands, the Gamilaraay people, and pay our respects to elders past, present and emerging. 
LRC \& ADM acknowledge support from an ARC Future Fellowship (FT160100206).
For the purpose of open access, the author has applied a Creative Commons Attribution (CC BY) licence to any Author Accepted Manuscript version arising
from this submission.

\section*{Data availability}
Underlying data will be shared on reasonable request to the authors.

%%%%%%%%%%%%%%%%%%%%%%%%%%%%%%%%%%%%%%%%%%%%%%%%%%



\bibliographystyle{mnras}
\bibliography{SF1letter} % if your bibtex file is called example.bib

\begin{thebibliography}{}
\makeatletter
\relax
\def\mn@urlcharsother{\let\do\@makeother \do\$\do\&\do\#\do\^\do\_\do\%\do\~}
\def\mn@doi{\begingroup\mn@urlcharsother \@ifnextchar [ {\mn@doi@}
  {\mn@doi@[]}}
\def\mn@doi@[#1]#2{\def\@tempa{#1}\ifx\@tempa\@empty \href
  {http://dx.doi.org/#2} {doi:#2}\else \href {http://dx.doi.org/#2} {#1}\fi
  \endgroup}
\def\mn@eprint#1#2{\mn@eprint@#1:#2::\@nil}
\def\mn@eprint@arXiv#1{\href {http://arxiv.org/abs/#1} {{\tt arXiv:#1}}}
\def\mn@eprint@dblp#1{\href {http://dblp.uni-trier.de/rec/bibtex/#1.xml}
  {dblp:#1}}
\def\mn@eprint@#1:#2:#3:#4\@nil{\def\@tempa {#1}\def\@tempb {#2}\def\@tempc
  {#3}\ifx \@tempc \@empty \let \@tempc \@tempb \let \@tempb \@tempa \fi \ifx
  \@tempb \@empty \def\@tempb {arXiv}\fi \@ifundefined
  {mn@eprint@\@tempb}{\@tempb:\@tempc}{\expandafter \expandafter \csname
  mn@eprint@\@tempb\endcsname \expandafter{\@tempc}}}

\bibitem[\protect\citeauthoryear{Belokurov \& Erkal}{Belokurov \&
  Erkal}{2019}]{belokurovCloudsArms2019}
Belokurov V.,  Erkal D.,  2019, \mn@doi [MNRAS] {10.1093/mnrasl/sly178}, 482,
  L9

\bibitem[\protect\citeauthoryear{Belokurov, Erkal, Deason, Koposov, De~Angeli,
  Wyn~Evans, Fraternali  \& Mackey}{Belokurov
  et~al.}{2017}]{belokurovCloudsStreamsBridges2017}
Belokurov V.,  Erkal D.,  Deason A.~J.,  Koposov S.~E.,  De~Angeli F.,
  Wyn~Evans D.,  Fraternali F.,   Mackey D.,  2017, \mn@doi [MNRAS]
  {10.1093/mnras/stw3357}, 466, 4711

\bibitem[\protect\citeauthoryear{Besla, Kallivayalil, Hernquist, {van der
  Marel}, Cox  \& Kere{\v s}}{Besla et~al.}{2012}]{beslaRoleDwarfGalaxy2012a}
Besla G.,  Kallivayalil N.,  Hernquist L.,  {van der Marel} R.~P.,  Cox T.~J.,
   Kere{\v s} D.,  2012, \mn@doi [MNRAS] {10.1111/j.1365-2966.2012.20466.x},
  421, 2109

\bibitem[\protect\citeauthoryear{Bressan, Marigo, Girardi, Salasnich, Dal~Cero,
  Rubele  \& Nanni}{Bressan et~al.}{2012}]{bressanPARSECStellarTracks2012}
Bressan A.,  Marigo P.,  Girardi L.,  Salasnich B.,  Dal~Cero C.,  Rubele S.,
  Nanni A.,  2012, \mn@doi [MNRAS] {10.1111/j.1365-2966.2012.21948.x}, 427, 127

\bibitem[\protect\citeauthoryear{Cheng et~al.,}{Cheng
  et~al.}{2022}]{chengKinematicalAnalysisSubstructure2022a}
Cheng X.,  et~al., 2022, \mn@doi [ApJ] {10.3847/1538-4357/ac5621}, 928, 95

\bibitem[\protect\citeauthoryear{Choi, Olsen, Besla, {van der Marel}, Zivick,
  Kallivayalil  \& Nidever}{Choi et~al.}{2022}]{choiRecentLMCSMC2022a}
Choi Y.,  Olsen K. A.~G.,  Besla G.,  {van der Marel} R.~P.,  Zivick P.,
  Kallivayalil N.,   Nidever D.~L.,  2022, \mn@doi [ApJ]
  {10.3847/1538-4357/ac4e90}, 927, 153

\bibitem[\protect\citeauthoryear{Cioni}{Cioni}{2009}]{cioniMetallicityGradientTracer2009}
Cioni M.-R.~L.,  2009, \mn@doi [A\&A] {10.1051/0004-6361/200912138}, 506, 1137

\bibitem[\protect\citeauthoryear{Cioni et~al.,}{Cioni
  et~al.}{2011}]{cioniVMCSurveyStrategy2011}
Cioni M.-R.,  et~al., 2011, \mn@doi [A\&A] {10.1051/0004-6361/201016137}, 527,
  A116

\bibitem[\protect\citeauthoryear{Cullinane et~al.,}{Cullinane
  et~al.}{2020}]{C20}
Cullinane L.~R.,  et~al., 2020, \mn@doi [MNRAS] {10.1093/mnras/staa2048}, 497,
  3055

\bibitem[\protect\citeauthoryear{Cullinane, Mackey, Da~Costa, Erkal, Koposov
  \& Belokurov}{Cullinane et~al.}{2021}]{C21}
Cullinane L.~R.,  Mackey A.~D.,  Da~Costa G.~S.,  Erkal D.,  Koposov S.~E.,
  Belokurov V.,  2021, \mn@doi [MNRAS] {10.1093/mnras/stab3350}, 510, 445

\bibitem[\protect\citeauthoryear{Cullinane, Mackey, Da~Costa, Erkal, Koposov
  \& Belokurov}{Cullinane et~al.}{2022}]{C22}
Cullinane L.~R.,  Mackey A.~D.,  Da~Costa G.~S.,  Erkal D.,  Koposov S.~E.,
  Belokurov V.,  2022, \mn@doi [MNRAS] {10.1093/mnras/stac733}, 512, 4798

\bibitem[\protect\citeauthoryear{De~Leo, Carrera, No{\"e}l, Read, Erkal  \&
  Gallart}{De~Leo et~al.}{2020}]{deleoRevealingTidalScars2020a}
De~Leo M.,  Carrera R.,  No{\"e}l N. E.~D.,  Read J.~I.,  Erkal D.,   Gallart
  C.,  2020, \mn@doi [MNRAS] {10.1093/mnras/staa1122}, 495, 98

\bibitem[\protect\citeauthoryear{Dobbie, Cole, Subramaniam  \& Keller}{Dobbie
  et~al.}{2014}]{dobbieRedGiantsSmall2014a}
Dobbie P.~D.,  Cole A.~A.,  Subramaniam A.,   Keller S.,  2014, \mn@doi [MNRAS]
  {10.1093/mnras/stu926}, 442, 1680

\bibitem[\protect\citeauthoryear{El~Youssoufi et~al.,}{El~Youssoufi
  et~al.}{2021}]{youssoufiStellarSubstructuresPeriphery2021}
El~Youssoufi D.,  et~al., 2021, \mn@doi [MNRAS] {10.1093/mnras/stab1075}, 505,
  2020

\bibitem[\protect\citeauthoryear{{Gaia Collaboration} et~al.,}{{Gaia
  Collaboration} et~al.}{2018}]{gaiacollaborationGaiaDataRelease2018b}
{Gaia Collaboration} et~al., 2018, \mn@doi [A\&A]
  {10.1051/0004-6361/201832698}, 616, A12

\bibitem[\protect\citeauthoryear{{Gaia Collaboration} et~al.,}{{Gaia
  Collaboration} et~al.}{2021}]{gaiacollaborationGaiaEarlyData2021b}
{Gaia Collaboration} et~al., 2021, \mn@doi [A\&A]
  {10.1051/0004-6361/202039657}, 649, A1

\bibitem[\protect\citeauthoryear{Girardi \& Salaris}{Girardi \&
  Salaris}{2001}]{girardiPopulationEffectsRed2001}
Girardi L.,  Salaris M.,  2001, \mn@doi [MNRAS]
  {10.1046/j.1365-8711.2001.04084.x}, 323, 109

\bibitem[\protect\citeauthoryear{Graczyk et~al.,}{Graczyk
  et~al.}{2020}]{graczykDistanceDeterminationSmall2020a}
Graczyk D.,  et~al., 2020, \mn@doi [ApJ] {10.3847/1538-4357/abbb2b}, 904, 13

\bibitem[\protect\citeauthoryear{Grady, Belokurov  \& Evans}{Grady
  et~al.}{2021}]{gradyMagellanicMayhemMetallicities2021}
Grady J.,  Belokurov V.,   Evans N.~W.,  2021, \mn@doi [ApJ]
  {10.3847/1538-4357/abd4e4}, 909, 150

\bibitem[\protect\citeauthoryear{Hatzidimitriou \& Hawkins}{Hatzidimitriou \&
  Hawkins}{1989}]{hatzidimitriouStellarPopulationsLargescale1989}
Hatzidimitriou D.,  Hawkins M. R.~S.,  1989, \mn@doi [MNRAS]
  {10.1093/mnras/241.4.667}, 241, 667

\bibitem[\protect\citeauthoryear{James et~al.,}{James
  et~al.}{2021}]{jamesPresenceRedGiant2021}
James D.,  et~al., 2021, \mn@doi [MNRAS] {10.1093/mnras/stab2873}, 508, 5854

\bibitem[\protect\citeauthoryear{Kirby, Cohen, Guhathakurta, Cheng, Bullock  \&
  Gallazzi}{Kirby et~al.}{2013}]{kirbyUniversalStellarMassStellar2013}
Kirby E.~N.,  Cohen J.~G.,  Guhathakurta P.,  Cheng L.,  Bullock J.~S.,
  Gallazzi A.,  2013, \mn@doi [ApJ] {10.1088/0004-637X/779/2/102}, 779, 102

\bibitem[\protect\citeauthoryear{Lewis et~al.,}{Lewis
  et~al.}{2002}]{lewisAngloAustralianObservatory2dF2002}
Lewis I.~J.,  et~al., 2002, \mn@doi [MNRAS] {10.1046/j.1365-8711.2002.05333.x},
  333, 279

\bibitem[\protect\citeauthoryear{Lucchini, D'Onghia  \& Fox}{Lucchini
  et~al.}{2021}]{lucchiniMagellanicStream202021}
Lucchini S.,  D'Onghia E.,   Fox A.~J.,  2021, \mn@doi [ApJL]
  {10.3847/2041-8213/ac3338}, 921, L36

\bibitem[\protect\citeauthoryear{Mackey, Koposov, Da~Costa, Belokurov, Erkal
  \& Kuzma}{Mackey et~al.}{2018}]{mackeySubstructuresTidalDistortions2018}
Mackey D.,  Koposov S.~E.,  Da~Costa G.,  Belokurov V.,  Erkal D.,   Kuzma P.,
  2018, \mn@doi [ApJ] {10.3847/2041-8213/aac175}, 858, L21

\bibitem[\protect\citeauthoryear{Majewski, Nidever, Mu{\~n}oz, Patterson,
  Kunkel  \& Carlin}{Majewski
  et~al.}{2008}]{majewskiDiscoveryExtendedHalolike2008a}
Majewski S.~R.,  Nidever D.~L.,  Mu{\~n}oz R.~R.,  Patterson R.~J.,  Kunkel
  W.~E.,   Carlin J.~L.,  2008, \mn@doi [Proc. IAU]
  {10.1017/S1743921308028251}, 4, 51

\bibitem[\protect\citeauthoryear{Massana et~al.,}{Massana
  et~al.}{2020}]{massanaSMASHingLowSurface2020}
Massana P.,  et~al., 2020, \mn@doi [MNRAS] {10.1093/mnras/staa2451}, 498, 1034

\bibitem[\protect\citeauthoryear{Nidever, Monachesi, Bell, Majewski, Mu{\~n}oz
  \& Beaton}{Nidever et~al.}{2013}]{nideverTidallyStrippedStellar2013}
Nidever D.~L.,  Monachesi A.,  Bell E.~F.,  Majewski S.~R.,  Mu{\~n}oz R.~R.,
  Beaton R.~L.,  2013, \mn@doi [ApJ] {10.1088/0004-637X/779/2/145}, 779, 145

\bibitem[\protect\citeauthoryear{Nidever et~al.,}{Nidever
  et~al.}{2017}]{nideverSMASHSurveyMAgellanic2017}
Nidever D.~L.,  et~al., 2017, \mn@doi [MNRAS] {10.3847/1538-3881/aa8d1c}, 154,
  199

\bibitem[\protect\citeauthoryear{Omkumar et~al.,}{Omkumar
  et~al.}{2020}]{omkumarGaiaViewStellar2020}
Omkumar A.~O.,  et~al., 2020, \mn@doi [MNRAS] {10.1093/mnras/staa3085}, 500,
  2757

\bibitem[\protect\citeauthoryear{Pardy, D'Onghia  \& Fox}{Pardy
  et~al.}{2018}]{pardyModelsTidallyInduced2018a}
Pardy S.~A.,  D'Onghia E.,   Fox A.~J.,  2018, \mn@doi [ApJ]
  {10.3847/1538-4357/aab95b}, 857, 101

\bibitem[\protect\citeauthoryear{Pieres et~al.,}{Pieres
  et~al.}{2017}]{pieresStellarOverdensityAssociated2017a}
Pieres A.,  et~al., 2017, \mn@doi [MNRAS] {10.1093/mnras/stx507}, 468, 1349

\bibitem[\protect\citeauthoryear{Pietrzy{\'n}ski et~al.,}{Pietrzy{\'n}ski
  et~al.}{2019}]{pietrzynskiDistanceLargeMagellanic2019}
Pietrzy{\'n}ski G.,  et~al., 2019, \mn@doi [Nature]
  {10.1038/s41586-019-0999-4}, 567, 200

\bibitem[\protect\citeauthoryear{Schwarz}{Schwarz}{1978}]{schwarzEstimatingDimensionModel1978}
Schwarz G.,  1978, \mn@doi [Ann. Statistics] {10.1214/aos/1176344136}, 6

\bibitem[\protect\citeauthoryear{Scowcroft, Freedman, Madore, Monson, Persson,
  Rich, Seibert  \& Rigby}{Scowcroft
  et~al.}{2016}]{scowcroftCarnegieHubbleProgram2016}
Scowcroft V.,  Freedman W.~L.,  Madore B.~F.,  Monson A.,  Persson S.~E.,  Rich
  J.,  Seibert M.,   Rigby J.~R.,  2016, \mn@doi [ApJ]
  {10.3847/0004-637X/816/2/49}, 816, 49

\bibitem[\protect\citeauthoryear{Sharp \& Birchall}{Sharp \&
  Birchall}{2010}]{sharpOptimalExtractionFibre2010}
Sharp R.,  Birchall M.~N.,  2010, \mn@doi [PASA] {10.1071/AS08001}, 27, 91

\bibitem[\protect\citeauthoryear{Subramanian et~al.,}{Subramanian
  et~al.}{2017}]{subramanianVMCSurveyXXIV2017}
Subramanian S.,  et~al., 2017, \mn@doi [MNRAS] {10.1093/mnras/stx205}, 467,
  2980
  
\bibitem[\protect\citeauthoryear{{van der Marel} \& Kallivayalil}{{van der
Marel} \& Kallivayalil}{2014}]{vandermarelThirdEpochMagellanicCloud2014}
{van der Marel} R.~P.,  Kallivayalil N.,  2014, \mn@doi [ApJ]
{10.1088/0004-637X/781/2/121}, 781, 121

\bibitem[\protect\citeauthoryear{Zivick et~al.,}{Zivick
  et~al.}{2018}]{zivickProperMotionField2018b}
Zivick P.,  et~al., 2018, \mn@doi [ApJ] {10.3847/1538-4357/aad4b0}, 864, 55

\bibitem[\protect\citeauthoryear{Zivick, Kallivayalil  \& {van der
  Marel}}{Zivick et~al.}{2021}]{zivickDecipheringKinematicStructure2021}
Zivick P.,  Kallivayalil N.,   {van der Marel} R.~P.,  2021, \mn@doi [ApJ]
  {10.3847/1538-4357/abe1bb}, 910, 36

\makeatother
\end{thebibliography}

%%%%%%%%%%%%%%%%%%%%%%%%%%%%%%%%%%%%%%%%%%%%%%%%%%

%%%%%%%%%%%%%%%%% APPENDICES %%%%%%%%%%%%%%%%%%%%%

%%%%%%%%%%%%%%%%%%%%%%%%%%%%%%%%%%%%%%%%%%%%%%%%%%

% Don't change these lines
\bsp	% typesetting comment
\label{lastpage}
\end{document}